\documentclass[12pt]{article}

\title{On the importance of grain boundaries in $HT_c$ films:
simulation results}

\author{R. Mulet and E. Altshuler\\
 Superconductivity Laboratory, Physics Faculty-IMRE,\\
University of Havana\\
10400 La Habana, Cuba}

\date{\today}

\begin{document}
\maketitle

\begin{abstract}
The importance of the distribution and ``weakness'' of grain boundary
junctions in the magnetic field dependence of the transport critical current
in $HT_{c}$ films is assessed through simulations.
The system is studied with the applied field either parallel or perpendicular
to the $\vec{c}$ axis of the sample.
For {\em realistic} sample parameters, it is demonstrated that the presence of
``high'' misorientation angles between grains depresses the zero - field
critical current density in both orientations, and provokes a transition from
pinning-mediated to Fraunhoffer-like field dependencies of the critical
current density. Our results also suggest that there is a threshold
misorientation angle above which the critical current density remains
constant. 


\end{abstract}

\section{Introduction}

One of the limiting factors found in the race to get large scale applications
from $HT_{c}$ superconductors 
is their granular character which constitutes a severe handicap to obtain
large values of critical current densities.

From today's variety of $HT_{c}$ superconducting materials,
 bismuth tapes seem to be the best
solution when we are looking for good mechanical properties 
and low cost performance figures (see for example \cite{Rieger_98} and 
\cite{Malozemoff_www}).
However, the highest values of critical current densities are obtained 
in $YBaCuO$ thin films,  so this material
 cannot be discarded as a possibility for current carrying applications
\cite{Hawsey_96}.

It is generally observed that
these thin films grow through a nucleation process resulting in ``islands'' or
``columnar grains''
\cite{Moeckly_94,Raistrick}.
The existence and nature of weak links 
between these columnar grains have been the subject of 
debate for years since they are  
strongly dependent on the 
deposition technique, the substrate, and deposition parameters.
However, there is a general agreement 
that in  thin films, contrary to what happens in ceramic
 superconductors, transport properties are dominated by pinning mechanisms
 instead of Josephson effects.
 
In spite of this general belief, 
a great fraction of  the published data in the field shows
 the presence of low and large angle boundaries between grains in $HT_c$
films,
(particularly polycrystalline), and at 
the same time, careful studies of Chaudhary et al \cite{Chaudary_88} and 
Gross \cite{Gross_93}
demonstrated that high angle grain 
boundaries drastically reduce the critical current of the junctions. 
To clarify the role of those boundaries, and of the pinning centers
 on the resulting critical 
current density,
 we developed a simple model for the 
transport properties of thin film superconductors. 

\section{The Model}

The critical current density within each superconducting grain 
is assumed to change with the field following a Kim-like model 
\cite{Kim} as:

\begin{equation}
J_{ci} \sim \frac{1}{1+H/H_o}	\label{eq:Jc_granos_O}
\end{equation}

\noindent where the subscript $ci$ stands for the different directions
of the current ($x,y,z$), $H$ is the magnetic field applied to the
sample, and $H_o$ is a parameter to be determined experimentally.

This equation can be writen in the following more convenient form for
computational purposes:

\begin{equation}
J_{ci} \sim \frac{1}{1+\beta p}	\label{eq:Jc_granos}
\end{equation}

\noindent where $p$ represents a normalized magnetic field, and
$\beta$ is a parameter which depends on the relation
between the field orientation and the sample axis.
 Following figure 1, the values of $\beta$ considered were:
if $\vec{H}//\vec{c}$, $\beta=1$ for $J_{cx}$ and $J_{cz}$ 
and $\beta=0$ for $J_{cy}$; if $\vec{H}\perp \vec{c}$
 ($\vec{H}$ along $\vec{x}$), $\beta=0.1$ for $J_{cz}$,
 $\beta=1.0$ for
$J_{cy}$  and $\beta=0$ for $J_{cx}$.
 To choose these values, we assumed that the intragranular current 
was depressed only by magnetic fields perpendicular to the current flow,
 and that the intrinsic pinning (i.e. that acting on the vortices 
lying parallel to the $ab$ planes when forced to move perpendicular to
 them) was an order of magnitude
 stronger than other sources of pinning \cite{Blatter_94, Sutton_94}. 

Up to this point, the model describes the field dependance of 
the critical current density in an {\it homogeneous}
medium (i.e., not {\it weak} links between grains).
However, if between the grains of the thin film high angle
tilt boundaries exist, the problem becomes more complicated. In fact, if
between two grains a high angle tilt boundary exists, the 
{\it intergranular} critical
current density follows the well-known
Fraunhoffer patern for a short Josephson junction \cite{Barone_82}:

\begin{equation}
	A \frac{\sin(\pi \frac{\Phi}{\Phi_o})}
		{\pi \frac{\Phi}{\Phi_o}}  \label{eq:Jc_junturas_O}
\end{equation}

\noindent where the prefactor $A$ depends on the angle boundary,
$\Phi$ is the magnetic flux at the junction, and $\Phi_o$ is the
flux quantum.

Now, we can rewrite (\ref{eq:Jc_junturas_O}) as a function of
$p=H/H_o$, and, after a straightforward algebra, it is transformed in:

\begin{equation}
	A \frac{\sin(\alpha p)}{\alpha p}  \label{eq:Jc_junturas}
\end{equation}

\noindent where $\alpha=\pi \frac{d \lambda H_o}{\Phi_o}$ 
for $\vec{H}//\vec{c}$ and $\alpha=\pi \frac{L \lambda H_o}{\Phi_o}$ 
for $\vec{H} \perp \vec{c}$, ($d$ and $L$ are represented in
figure 1).  

A granular thin film can not be modeled as a pure ceramic
superconductor since, having just a small fraction of high angle tilt
boundaries, we can find paths of high critical current densities were
its dependence with the applied field is determined by the equation 
(\ref{eq:Jc_granos}). So, 
to model the system  we proposed a tridimensional array of grains, 
a fraction $q$
of them was strongly coupled (which means with misorientation angle
smaller than $7^o$ occur between grains \cite{Moeckly_94})
, while a fraction $(1-q)$ is coupled through weak links produced by
high angles between grains. For the strongly coupled grains the
critical current density was detertermined by means of
(\ref{eq:Jc_granos}), while for the weakly coupled grains equation
(\ref{eq:Jc_junturas}) was used.

The calculation of the critical current density of the system was
based on the Minimum Cut Algorithm
already used to solved similar problems \cite{Mulet_97, Chvatal}. This
algorithm allows the calculation of the maximum flow in a random
system. It is basically composed by two operations. A first one to
determine the paths on the system where the current can flow, and a
second one to augment the flow (of current) at the bonds (our links
between grains). Once it is impossible to find a new path or to augment
the flow through the already found paths, we say we obtained the
maximum current (critical current) of the system.

We choose for our simulations
$\alpha=20$ for $\vec{H}//\vec{c}$ which corresponds to 
$L=600nm$ and $\lambda=30nm$, and $\alpha=400$ for $\vec{H}\perp \vec{c}$
 corresponding to
$d=200nm$ and $\lambda=200nm$ as typically reported for $YBaCuO$ films
(see figure1) \cite{Moeckly_94,Raistrick}.
 Kim's characteristic field $H_o=1T$ was assumed. The possibility of
modeling granularity in such anisotropic fashion (but only in the
light of a simple parallel ensemble of Josephson junctions) has been
suggested earlier by Altshuler et al \cite{Altshuler_95}. We used
systems of dimensions $16 \times 16 \times 16$
(which mimics a $\sim 10\times10 \mu m^2$ bridge performed on a thin film with
average grain diameter of $600 nm$)
and averaged over 10 different configurations to improved the
statistics. The parameter $p$ was always varied between 0 and 3.

One more remark about the different configurations used is needed. 
If the applied field 
 points parallel to the $\vec{c}$
direction, junctions perpendicular to the $ab$ plane (shaded in figure 1)
are affected by equation (\ref{eq:Jc_junturas}),
while, if it is applied parallel to the $ab$ plane, junctions lying in that
plane are not affected.

\section{Results}

To approach real values of $YBaCuO$ films we extracted $A$ and $q$
from the
combination of two experimental results: the statistics of
 boundaries angles 
in a $YBaCuO$ films reported by \cite{Moeckly_94}, and the 
angle dependence of 
the critical current density measured by Ivanov et 
al \cite{Ivanov_91} on $YBaCuO$ 
bycristals.
Figure 2 shows the simulated field dependence of the critical current 
density corresponding to films with different microstructures. The upper 
curve represents the critical current versus field dependence of a system 
without weak-links, while the lower contains a $50\%$ of high angle 
boundaries.
 $80\%$ and $20\%$ of these boundaries reduce the critical current by 
approximate factors 
of 2 and 100 respectively, corresponding to misorientation angles of
$8^o$ and $20^o$ \cite{Moeckly_94,Ivanov_91}. As observed in 
figure 2 the introduction of weak-links provokes 
a reduction of the zero field critical current density by a factor $0.67$. 
It also induces a steeper $J_{c}(p)$ characteristic in the ``low field'' 
region, and several maxima which suggests a superposition of Josephson 
patterns, as observed in $YBaCuO$ polycrystals with a small number of 
grains \cite{Oscar_95,Oscar_99}.

To model a  better sample, we calculated $J_c$ for a system
without boundaries reducing the critical current  by a 
factor 100. The results are shown in figure 3, were the critical current 
density as a function of $p$ is plotted for a set of systems with $q$ 
ranging from $0.50$ to $1.0$. Even 
in this case the critical current density depends on the number of 
weak-links.
However, interestingly enough, the suppression of the ``worst'' boundaries 
doesn't improve significantly the critical current density at zero field. 
This suggests that the number of weak-links is more important than 
their ``weakness'' regarding the absolute values of the critical current.

To further explore this idea
 in figure 4, we compared the critical current dependencies with the 
applied field of samples with different kinds of weak-links and $q=0.5$. 
The upper curve represents angles in the range $7^o-10^o$
, i.e. reducing $J_{c}$ by a factor of 2, while the remaining curves 
represent angles of $12 ^o, 15^o, 25^{o}$ and $40^{o}$
 which roughly 
reduce the critical current by 4, 10, 
100 and 10000 respectively \cite{Moeckly_94,Ivanov_91}. The figure shows  again similar  $J_{c}$
 vs $p$ dependencies for high values of the angles. For high fields all
 the curves behave in the same manner, indicating that, in this regime,
 the number of weak links is more important than their quality. 
However, for zero field, different values of critical current densities
 are obtained depending on the misorientation angle between grains. 
Figure 5 shows this dependence, demonstrating
 that angles greater than $25^o$, do not considerably change the critical 
current densities values, while a strong improvement in $J_c$ can 
be obtained by diminishing the angle below $25^o$, which is qualitatively 
coherent with the data of Wu et al \cite{Wu_95}.

Figure 6 shows the field dependence of the critical current density for the
$\vec{H} \perp \vec{c}$ configuration. When compared with the figure 3, 
the effect of an anisotropic set of parameters is revealed in a less 
strong 
field dependence of $J_{c}$. However, the depression in the zero field
 critical
current density is roughly similar to the $\vec{H}//\vec{c}$ configuration. 
Here the weak-linked component also introduces a steep $J_{c}(p)$
 characteristics in the ``low field'' region, while Josephson 
assembly-like maxima are observed as $q$ decreased.

\section{Conclusions}
From these results we conclude that the presence of high angle boundaries 
 reduces the critical current density of  superconducting thin films
and provokes a transition form pinning-mediated to Fraunhoffer-like patterns
in its magnetic field dependencies. Our results also suggest that
 the amplitude of the misorientation
angles between grains {\em does not} change the values of critical current
density for high values of the applied field, while for low applied field,
differences appear only for small angle values.

\section*{Acknowledgments}
We like to thank, O. D\'{\i}az, D. Bueno and H. Herrmann for collaboration
during the initial stage of the computer work. We are also indebted to S.
Garc\'{\i}a for the computer facilities provided and to O. Ar\'es and C. Hart
for useful comments and the revision of the manuscript. The paper was
also largely beneffited from the comments of the referee. This work was also 
partially supported by the University of Havana ``Alma Mater'' Program, 1999.

\newpage

\section*{Figure Captions}

\hspace{0.5cm} {\bf Figure 1} Diagram of the system used in our simulations
 
{\bf Figure 2} $J_c$ vs $p$ dependences for  $\vec{H} // \vec{c}$. Upper
curve: q=1, Bottom curve: q=0.5. Two qualities of high misorientation angles,
$A=0.5$ and $A=0.001$ 

{\bf Figure 3} $J_c$ vs $p$ dependences for $\vec{H} // \vec{c}$. From top to
bottom: $q = 1, 0.8, 0.7$ and $ 0.5$. One quality of high misorientation
angles,
$A=0.5$

{\bf Figure 4} $J_{c}$ vs $p$ dependences, $\vec{H} // \vec{c}$,
$q=0.5$. From top to bottom: $A=0.5, 0.25, 0.1, 0.01$ and $0.0001$
correspondig to misorientation angles of $7^o < \theta<10^o$, $15^o, 25^o$ and
$40^o$.

{\bf Figure 5} $J_c$ dependence with the approximate misorientation angle,
$q=0.5$

{\bf Figure 6} $J_c$ vs $p$ dependences, $\vec{H} \perp \vec{c}$. 
>From top to bottom $q=1, 0.8, 0.7$ and $ 0.5$

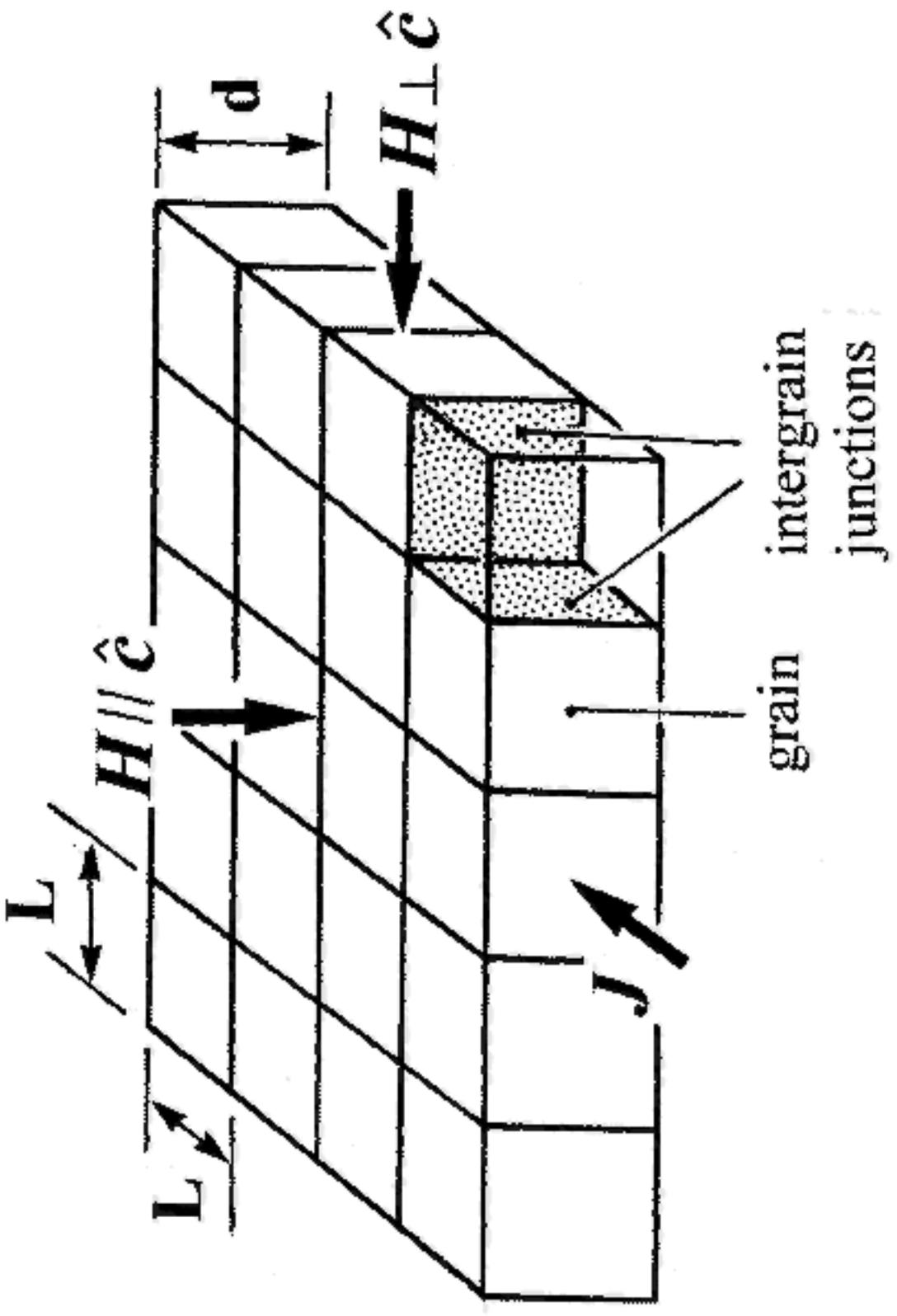
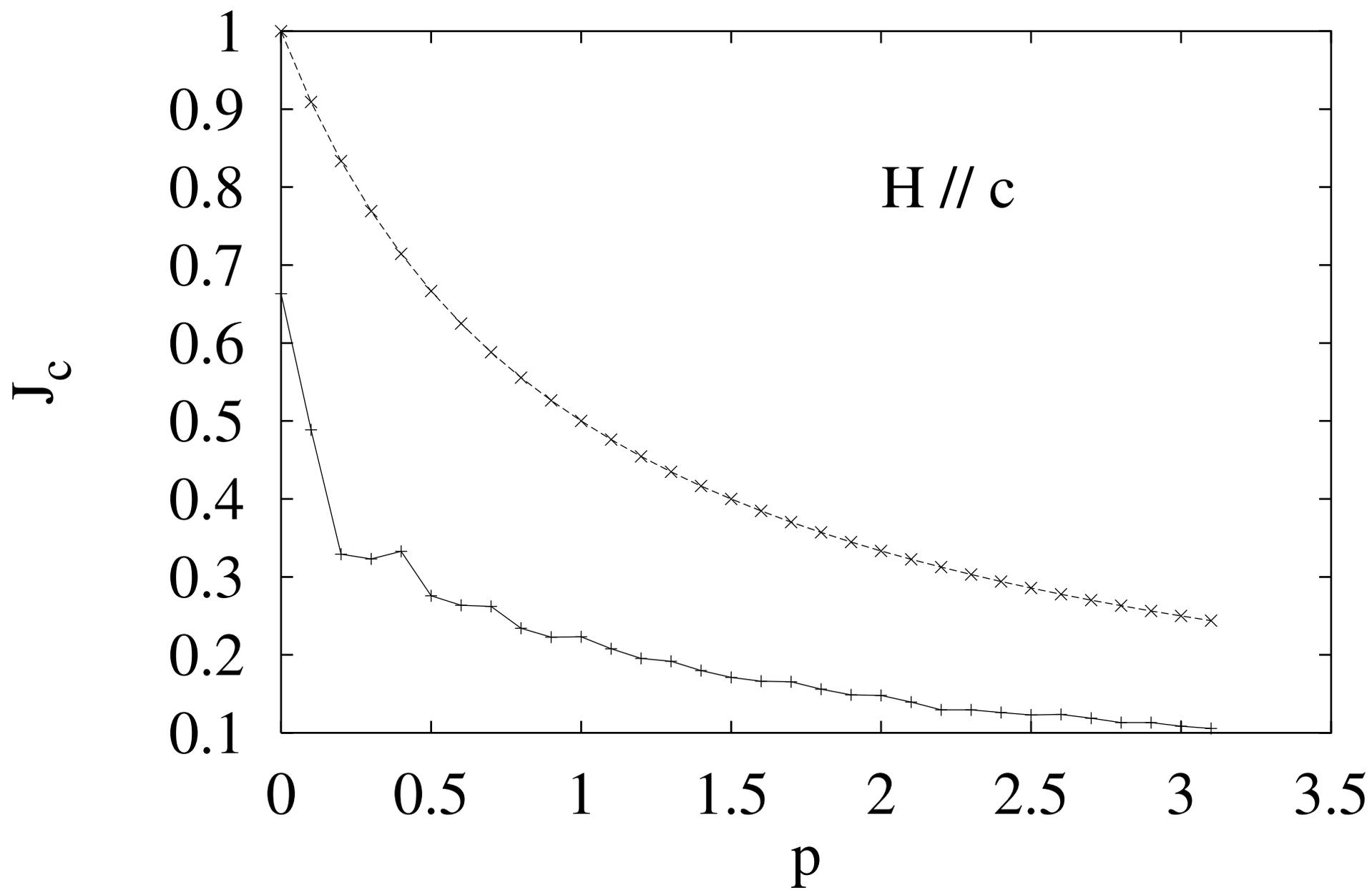
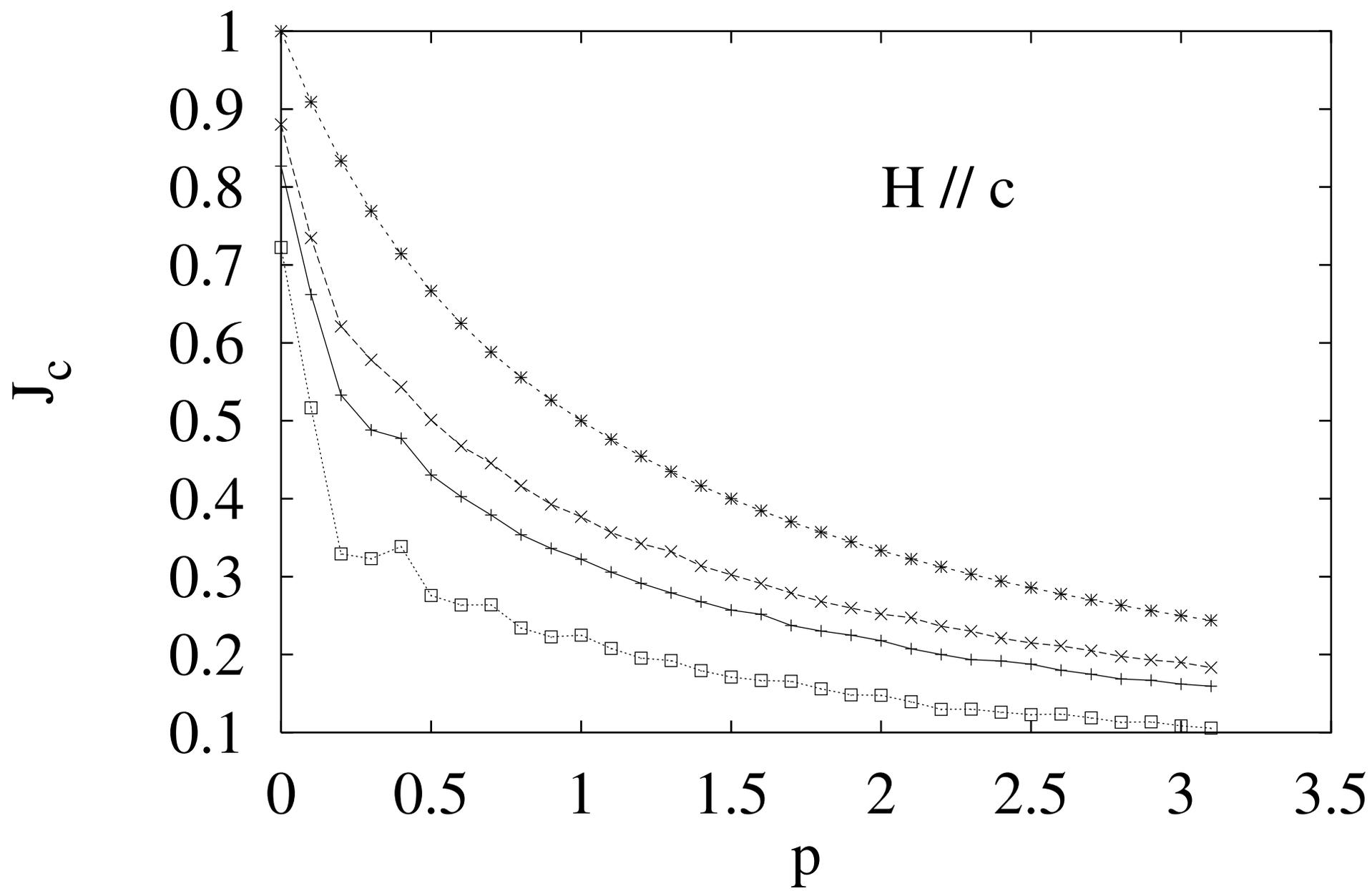
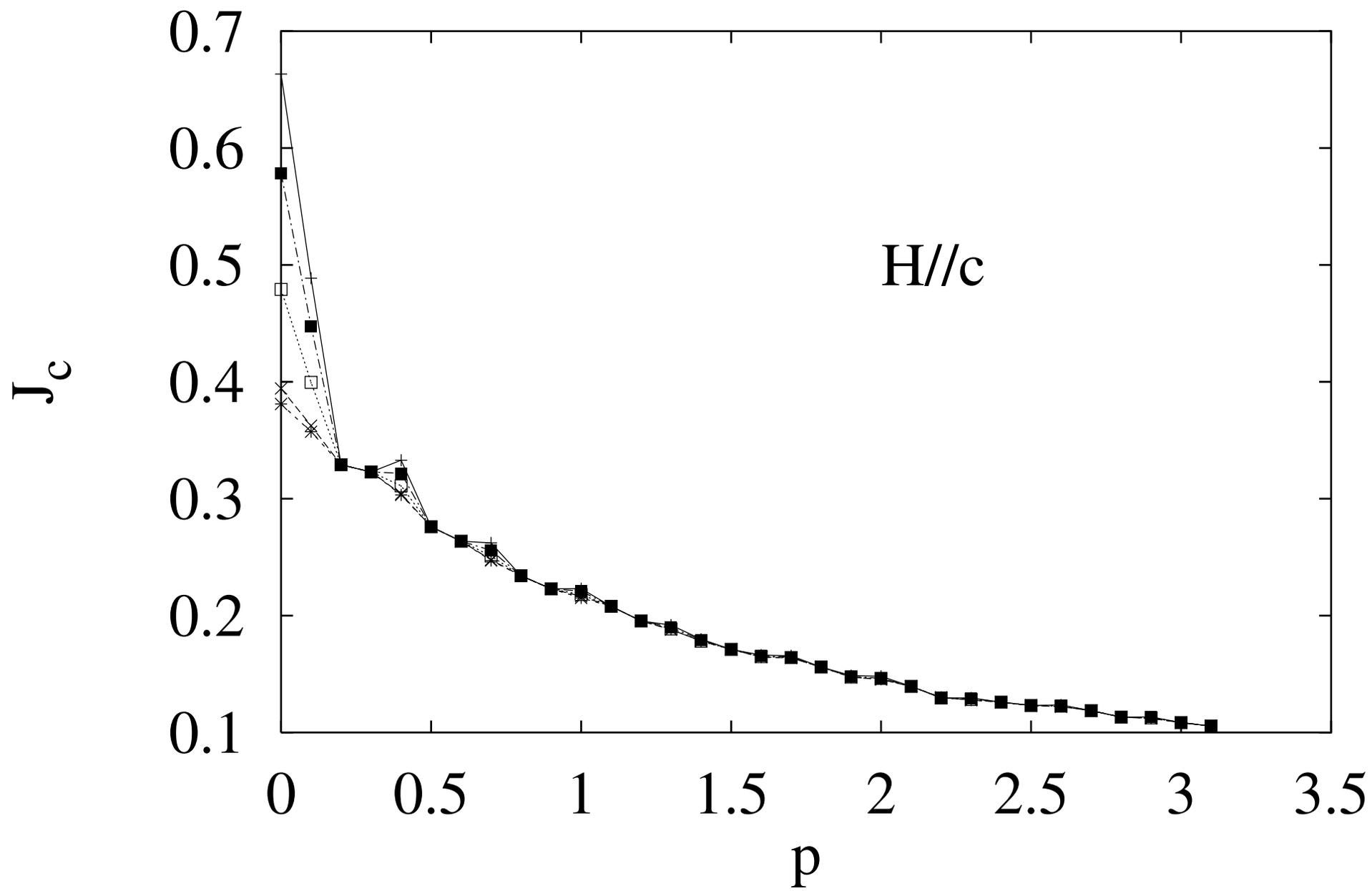
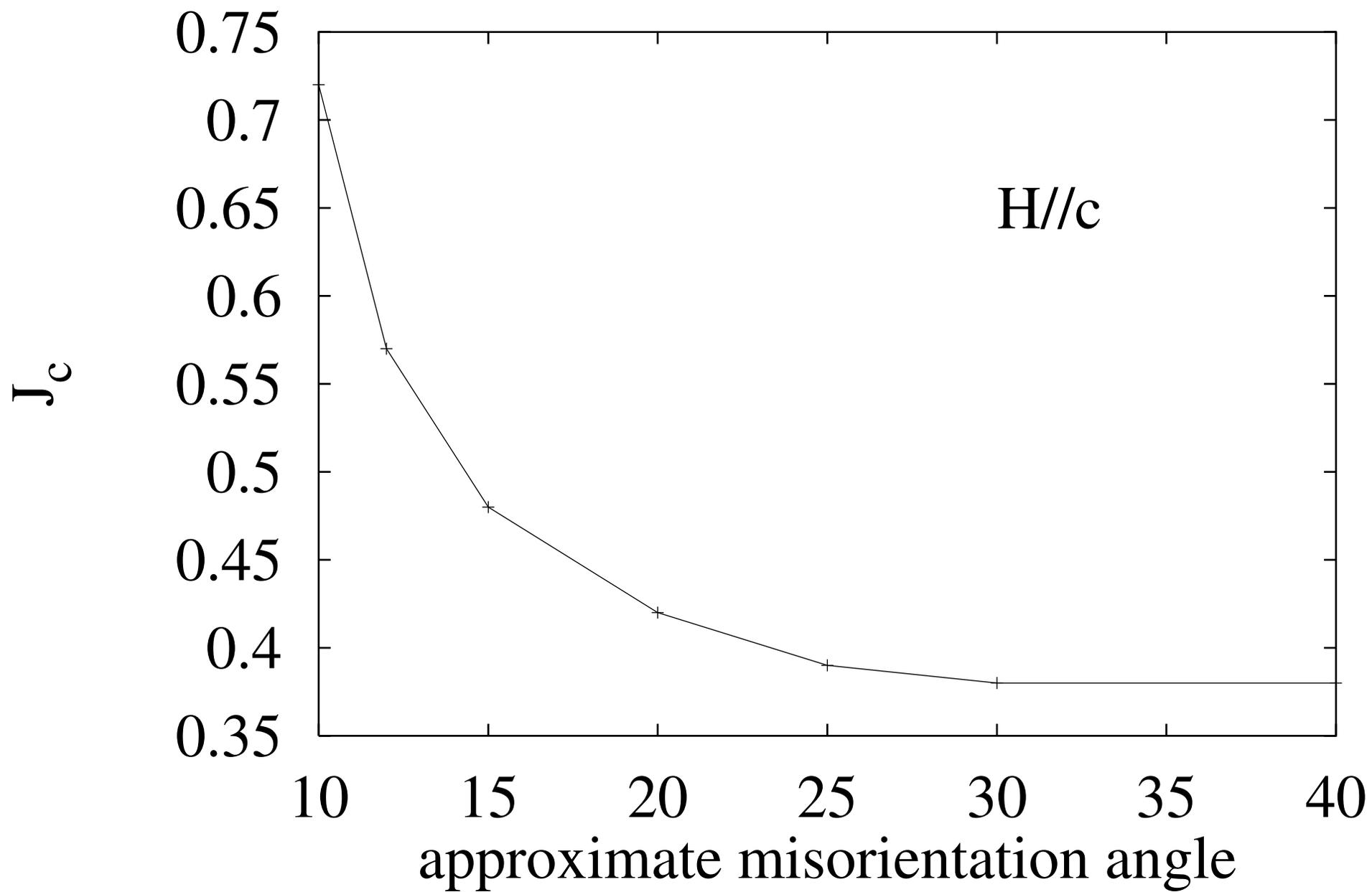
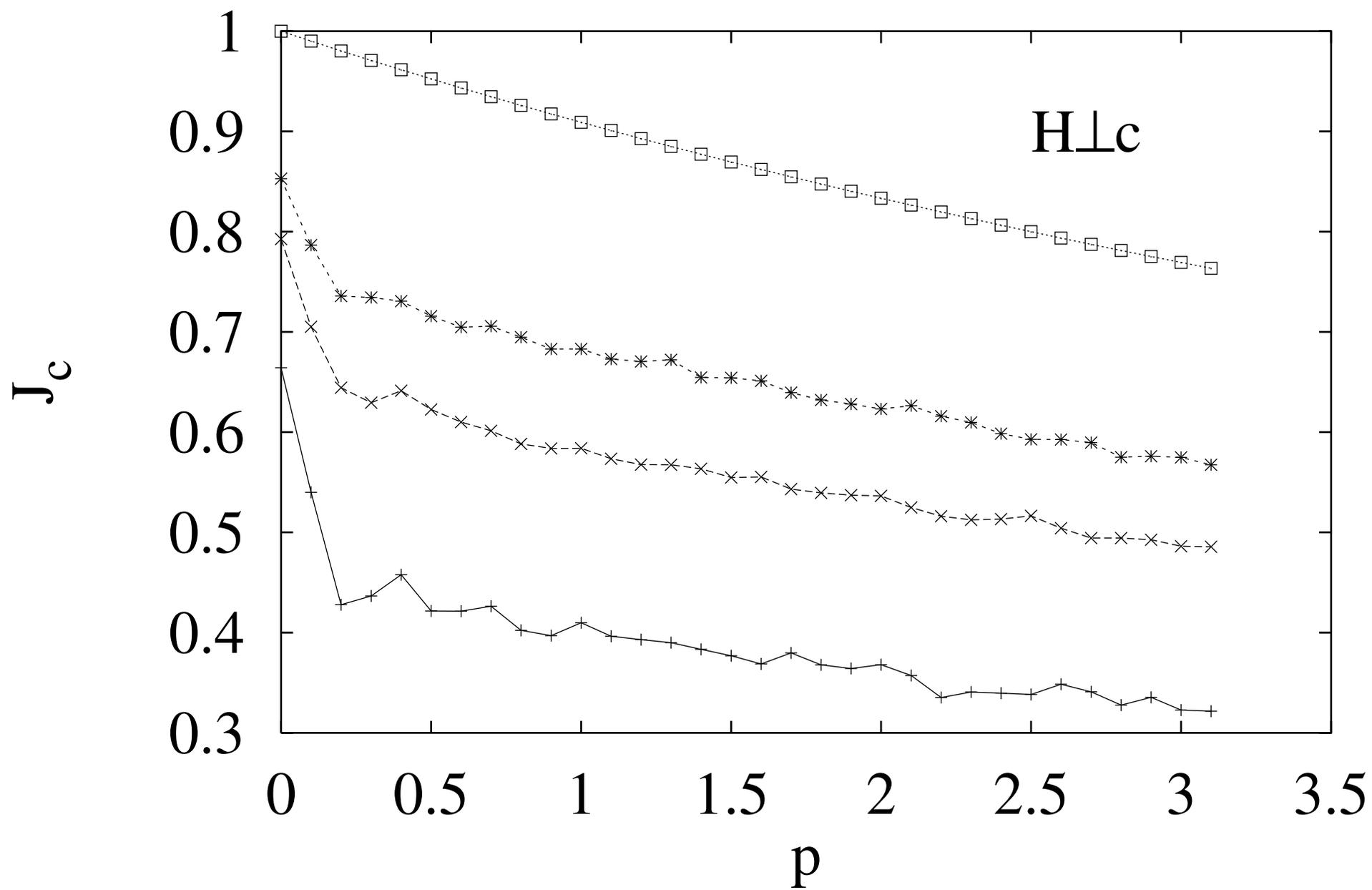
\end{document}